%% file: 3672.DASSM.tex
\font\caps=cmcsc10 at 12pt
\font\eightrm=cmr8 at 8pt
\documentclass{oxarticle}
\usepackage{amsmath, amssymb}
\input{Johnmacros15}

\newcommand{\articlenumber}{\LT 3672.DASSM}

\proofmodefalse
\begin{document}
\begin{center}

\vspace*{1in}
{  \huge Does SUSY know about the Standard Model? \\
 }

\vspace*{.1in}
%

     
\renewcommand{\thefootnote}{\fnsymbol{footnote}}

%
{\caps J. A. Dixon\footnote{jadixg@gmail.com}\\Toronto, Canada}\\[1cm] 

{\bf Abstract}
\end{center}
\normalsize
%

  The BRST cohomology of free chiral SUSY has a wealth of \EI s.  When one adds a
superpotential to the free theory, the extention of the \EI s leads to some constraints on that superpotential. 
A particularly simple solution of those constraints is based on a  $3 \times 3$ matrix of nine chiral superfields, and then the superpotential is simply the  determinant of that matrix.  

It is remarkable that this same theory is also a plausible  basic version of  the SUSY Standard Model for one Lepton family, and then the nine superfields are seen to be a left 
$SU(2)_{\rm Weak}$ Lepton Doublet, Two Higgs Doublets, a Right Electron Singlet, a Right Neutrino Singlet and a Higgs singlet.  Moreover, the algebra is consistent with the notion  that the other two observed Lepton families arise from the coupling of the 
\EI s.


  \section{A Simple Form for the Supersymmetric Standard Model}

\la{introsection}

 Why does the Standard Model contain such a strange array of irreducible representations of the gauge groups $SU(3) \times SU(2) \times U(1)$? Why are there three families of Quarks and Leptons, with Left $SU(2)_{\rm Weak}$ Doublets and Right Singlets?
 Most attempts to explain the Standard Model have focused on deriving it from the gauge theory of larger groups such as SU(5) \ci{SU5}. However, these attempts tend to make experimental predictions that appear to be  wrong \ci{ExptSU5}.  On the other hand, the Standard Model itself seems surprisingly  viable  from an experimental point of view \ci{SMexpmt}.

This paper discusses  a new route towards understanding the origin of the Standard Model. 
  The route arises from the BRST cohomology of its supersymmetric version \ci{SSM,SUSY09}, and it focuses, at first, on the Lepton and Higgs sectors alone. In other words, here we will look at the chiral superfields, and ignore the gauge superfields for the present\footnote{This is necessary because the gauge superfields present currently unsolved problems for their BRST cohomology and \EI s.}.

In \ci{YMextrainv}, it was demonstrated that a new way to look for new theories is to take a free theory and find its \EI s\footnote{ It was shown that \YM\ is `contained within' free gauge theory, in the form of an \EI.  One can obtain the full \YM\ by adding the \ELI\ to the theory and then completing the result so that it satisfies the 
\PB. This generates the Lie algebra structure constants  $f^{abc}$ and their Jacobi Identity, from the free theory, through the satisfaction of constraints that emerge from the BRST formalism.}.
In this paper we examine what happens if we look at the 
\EI s of free chiral SUSY, and attempt to extend them to a theory with a cubic superpotential. Surprisingly, one of the simplest possible non-trivial solutions for the resulting constraints is  a special version of the SUSY \SM, which we call the DASSM.

This simplified  `DASSM Action' for the Supersymmetric Standard 
Model (SSM) is a special version with only one Lepton family.  It can be written in the following very simple form:
\be
{\cal A} = 
\int d^4 x \;d^4 \q
\; {\rm Tr} \lt \{ A A^{\dag}
\rt \}
+
g_1 \int d^4 x \;d^2 \q
\;{\rm Det} \lt \{ A 
\rt \}
+
\og_1 \int d^4 x \;d^2 \oq
\;{\rm Det} \lt \{ A^{\dag} 
\rt \}
\la{simplest}
\ee
where $A^{IA}$ (where $I,A=1,2,3$) is a $3 \times 3$ matrix of chiral superfields, which means that its Determinant is a cubic term in the superfields, so that (\ref{simplest}) is a renormalizable action.  This  Determinant of this matrix can be written in several equivalent ways:
\be
{\rm Det}\; \{ A\} 
\equiv
|A| \equiv
\fr{1}{6}\ve_{IJK} \ve_{ABC} A^{IA} A^{JB} A^{KC}
\equiv \fr{1}{3} g_{MNP} A^M A^N A^P
\la{symtensormultsymmofsm234}
\ee

 To see that  (\ref{simplest}) is the 
action for  the  supersymmetric version of the Standard Model \ci{SSM}, for the special case where the matter content consists of one Lepton family, we write the components in the form:
\be
A^{IA} \ra 
\lt (
\begin{array}{ccc}
H^i & K^i & L^i\\
S & P
& J
\\
\end{array}
\rt )
\la{withoutcouplings}\ee
Then it is easy to show that we get the following familiar form for the superpotential polynomial:
\be
{\rm Det}
\lt \{
A \rt \}=
{\cal P} =  H^i  K_{i} J
+  K^i  L_{i} S
+   L^{i} H_{i} P
\la{basicPfromdet}
\ee

 The form (\ref{symtensormultsymmofsm234}) implies that we must make the contractions, and raise and lower indices,  with the two index antisymmetric tensors in two dimensions: 
\be  \ve^{ij}=- \ve^{ji}, H^i = \ve^{ij} H_j, \;{\rm etc.}
\ee
The form (\ref{basicPfromdet}) is the usual form of the superpotential for the SSM, except for the presence of a  right singlet neutrino S and a singlet Higgs J. These make all the difference of course, because they are needed to complete the matrix in equation 
(\ref{withoutcouplings}).

These $SU(2)_{\rm Weak}$ Singlet Superfields S and J are  not included in the `minimal' SSM, which is usually restricted to just the superfields L,H,K,P plus the Quarks.   Here $H^i$ and $K^i$ are Higgs $SU(2)_{\rm Weak}$
doublets. As is well known, these two doublets are needed to give rise to mass for the up and down quarks.  For the present case they are needed to give mass to the electron and the neutrino, as well.  $L^i$  is a left Lepton $SU(2)_{\rm Weak}$
doublet and $P$ is a right positron singlet.

The two extra superfields $S$ 
and $J$ are not excluded by experiment, and they are actually quite useful for the theory.
The superfield S gives rise to a Dirac mass for the neutrino, and the neutrino certainly has mass \ci{massneutrino}.  The Higgs singlet J can be used to give mass to the particles while it also spontaneously breaks the gauge symmetry.
This can be accomplished by adding the term $g' m^2 J$ to the superpotential.

The supermultiplet $A^{IA}$ appears in the superpotential (\ref{symtensormultsymmofsm234}) in a maximally symmetric way, with an invariance $SU(3) \times SU(3)$, corresponding to the
fact that $\ve_{IJK}$ and $\ve_{ABC}$ in
(\ref{symtensormultsymmofsm234})  are invariant tensors of the two  $SU(3)$ groups, with the transformation
\be
A^{JB} \ra M^{J}_{1,K} M^{B}_{2,C} A^{KC}
\;{\rm where} \;M_i\in SU(3)
\la{fatgroup}
\ee
This invariance is also evident from the Determinant form of the superpotential in
 (\ref{symtensormultsymmofsm234}), since 
\be
{\rm Det}
\lt \{
M^{J}_{1,K} M^{B}_{2,C} A^{KC}
\rt \}
=
{\rm Det}
\lt \{
M_1 \rt \}
{\rm Det}
\lt \{
A \rt \}
{\rm Det}
\lt \{
M_2 \rt \}
=
{\rm Det}
\lt \{
A \rt \}
\ee There are nine complex chiral superfields in this action, and there are sixteen real parameters in the above transformations.
Note that the kinetic term in (\ref{simplest}) is also invariant under these transformations, provided that they are either space-time constant or gauged\footnote{The form (\ref{symtensormultsymmofsm234})  implies that there is an SU(2) invariance associated with these $i$ indices. It turns out to be 
$SU(2)_{\rm Weak}$.
  The weak hypercharge is also included in the $SU(3) \times SU(3)$ invariance, as is Lepton number (and Baryon number for the Full DASSM). 
Here it is not clear how the masses of the electron, neutrino and Higgs are going to be different.  This will be discussed below in section
 \ref{basicliealg}.}.

Since this basic one-Lepton superpotential 
 (\ref{symtensormultsymmofsm234}) arises from `Double Antisymmetrization' (DA), which is equivalent to the determinant, as noted in (\ref{symtensormultsymmofsm234}), we will call it the `Basic DASSM' for clarity. 

It is simple to construct the corresponding action with three generations of Quarks and Leptons, and we will
call that the `Full DASSM'. This action is written down in section \ref{fulldassmsec} below.  However, it is possible that the Full DASSM might be misguided, since the other two families of Leptons may arise in a different way than just simple repetition\footnote{ They may arise from the couplings to the dotspinors, as explained below in section \ref{basicliealg}. This would work similarly for the Quarks, except that the first family of Quarks, with a colour $SU(3)$ index, does not seem to be required, which is admittedly very puzzling.  This means that the other two families of Quarks are not required either, for this model, as presently understood.}.

We also need to incorporate the fact that the 
SSM does not have the same mass for the electron and neutrino, which means that really the form 
\be
{\rm Det}
\lt \{
A' \rt \}=
{\cal P} =  g H^i  K_{i} J
+ s K^i  L_{i} S
+  p L^{i} H_{i} P
\la{basicPfromdetwithconstants}
\ee
is needed.  The symmetry is then present, but explicitly broken in this way.  But the symmetry is still there in the sense that it governs the kind of fields that are present, even though the couplings are varied. This can be obtained from
\be
A^{'IA} \ra 
\lt (
\begin{array}{ccc}
H^i & K^i & L^i\\
s S &pP
& g J
\\
\end{array}
\rt )
\la{withcouplings}\ee
Note that if we use this in the kinetic part, the normalizations of the fields are changed, which means that the symmetry is broken.
We shall return to this action and to the issue of the broken symmetry below in section \ref{basicliealg}. It turns out that for the \EI s to exist, the form (\ref{withcouplings}) works perfectly well.

\section{\EI s of Chiral SUSY}

\la{bootdassm}

In \ci{YMextrainv}, the \YM\ was shown to emerge from free gauge theory when one adds a certain \ELI\ to the action. This `seed' leads to the full \YM\ through the need to complete the action so that it satisfies 
the  \PB. In this paper we examine a rather similar situation which arises for chiral SUSY theories, and it leads us to the action (\ref{simplest}) above.

We start with the usual action 
 \ci{superfieldnotation,west,superspace,WB,
ferrarabook}:
\be
{\cal A}_{\rm Interacting\;SUSY}=
{\cal A}_{\rm Free}+
{\cal A}_{\rm Superpotential}
\la{fullsusyaction}\ee
where
\be
{\cal A}_{\rm Free}=
\int d^4 x d^4 \q
 {\widehat A}^i 
{\widehat \A}_i  
+
\int d^4 x d^2 \q
\lt \{ {\widehat \Lam}_i 
\lt (
 C^{\a} Q_{\a}  +
\oC^{\dot \a} \oQ_{\dot \a}
\rt ){\widehat A}^i  \rt\}
+
\int d^4 x d^2 \q
\lt \{ {\widehat {\ov \Lam}}^i 
\lt (
 C^{\a} Q_{\a}  +
\oC^{\dot \a} \oQ_{\dot \a}
\rt ){\widehat \A}_i  \rt\}
\la{chiralsusyaction}
\ee
and
\be
{\cal A}_{\rm Superpotential}
=
\fr{1}{3}
\int d^4 x d^2 \q
g_{ijk}
{\widehat A}^i 
{\widehat A}^j 
{\widehat A}^k 
+
\fr{1}{3}
\int d^4 x d^2 \oq
\og^{ijk}
{\widehat \A}_i 
{\widehat \A}_j 
{\widehat \A}_k 
\ee

and the \PB\ takes the form\footnote{Again here we have to perform some manipulation to remove exterior derivative complications.  This can be found in \ci{jumps} for example.}:
\be
{\cal P}_{\cal A}=\int d^4 x d^2 \q 
\lt \{
\fr{\d {\cal A}}{\d {\widehat \Lam}_i}
\fr{\d {\cal A}}{\d {\widehat A}^i}
\rt \}
+
\int d^4 x d^2 \oq 
\lt \{
\fr{\d {\cal A}}{\d {\widehat {\ov \Lambda}}^i}
\fr{\d {\cal A}}{\d {\widehat {\ov A}}_i}
\rt \}
\la{PBAchiral}
\ee
Here is the relevant BRST operator that we get from taking the `square root' of 
${\cal A}_{\rm Chiral\;SUSY}$ for the free case and the case where there is a superpotential:
\be
\d_{\rm Free}
=
\d_{\rm Field\;Equation\;Free}
+
\d_{\rm Zinn}
+
\d_{\rm Little}
\la{bigbrstsusy}
\ee
\be
\d_{\rm Interacting\;SUSY}
=
\d_{\rm Field\;Equation}
+
\d_{\rm Zinn}
+
\d_{\rm Little}
\la{bigbrstsusy}
\ee
where
\be
\d_{\rm Field\;Equation}
=\d_{\rm Field\;Equation\;Free}
+\d_{\rm Field\;Equation\;Superpotential}
\ee
and
\be
\d_{\rm Field\;Equation\;Free}
=
\int d^4 x \; d^2 \q\;
\ovD^2 
{\widehat \A}_i  
\fr{\d  }{\d {\widehat \Lambda}_i}  
+
\int d^4 x \; d^2 \oq\;
D^2 {\widehat A}^i  
 \fr{\d  }{\d {\widehat {\ov \Lambda}}^i}  
\la{fieldeqsusyfree}
\ee
\be
\d_{\rm Field\;Equation\;Superpotential}
=
\int d^4 x \; d^2 \q
\;
 g_{ijk} {\widehat A}^j {\widehat A}^k
\fr{\d  }{\d {\widehat \Lambda}_i}  
+
\int d^4 x \; d^2 \oq
\;  \og^{ijk} {\widehat \A}_j
 {\widehat \A}_k
 \fr{\d  }{\d {\widehat {\ov \Lambda}}^i}  
\la{fieldeqsusy}
\ee

and
\[
\d_{\rm Zinn}
=
\int d^4 x \; d^2 \q
\lt (
 C^{\a} Q_{\a} +
\oC^{\dot \a} \oQ_{\dot \a} 
\rt ) 
{\widehat \Lambda}_i
\fr{\d  }{\d {\widehat \Lambda}_i}  
\]
\be
+
\int d^4 x \; d^2 \oq
\lt (
 C^{\a} Q_{\a}  +
\oC^{\dot \a} \oQ_{\dot \a}
\rt ){\widehat {\ov \Lambda}}^i
 \fr{\d  }{\d {\widehat {\ov \Lambda}}^i}  
\la{zinnOthersusy}
\ee
and
\[
\d_{\rm Little}
=
\int d^4 x \; d^2 \q
\lt (
C^{\a} Q_{\a} +
\oC^{\dot \a} \oQ_{\dot \a} 
 \rt )
 {\widehat A}^i
\fr{\d   }{\d {\widehat A}^i}  
\]
\be
+
\int d^4 x \; d^2 \oq
\lt (
C^{\a} Q_{\a} +
\oC^{\dot \a} \oQ_{\dot \a} 
 \rt )
 {\widehat \A}_i
\fr{\d   }{\d {\widehat \A}_i}  
\la{BRSTlittlechiral}
\ee

Note that
\be
{\d}_{\rm Interacting\;SUSY}^2=\dF^2 
= C^{\a} \oC^{\dot \b} \pa_{\a \dot \b}
\equiv  0
\ee
where the equivalence applies to integrated local polynomials\footnote{We need to define superspace translation operators $ Q_{\a}$
$\oQ_{\dot \a}, 
\pa_{\a \dot \b}$ and  superspace chiral derivatives
 $ D_{\a}$
$\ov D_{\dot \a}$. These are made from  derivatives in terms of the ordinary variables of superspace, namely 
\be
\q_{\a}, \oq_{\dot \b},x_{\a \dot \b}
\la{explcitvariables}
\ee
and they take the form
\be
Q_{\a}= 
\fr{\pa}{\pa \q^{\a}}
+ \fr{1}{2}\oq^{\dot \b} \pa_{\a \dot \b}
\;;D_{\a}= 
\fr{\pa}{\pa \q^{\a}}
- \fr{1}{2}\oq^{\dot \b} \pa_{\a \dot \b}
\ee
\be
\oQ_{\dot \b}= \fr{\pa}{\pa \oq^{\dot \b}}
+ \fr{1}{2}\q^{\a} \pa_{\a \dot \b}
;\;
\ov D_{\dot \b}= \fr{\pa}{\pa \oq^{\dot \b}}
- \fr{1}{2}\q^{\a} \pa_{\a \dot \b}
\ee
Note that
\be
\lt \{
Q_{\a},
\oQ_{\dot \b}
\rt \}=
  \pa_{\a \dot \b}.
\ee}.

Sometimes it is useful to adopt the short notation:
\be
\d_{\rm SS} =
C^{\a} Q_{\a} +
\oC^{\dot \a} \oQ_{\dot \a}
\la{ssderivative}.
\ee
Note that
\be
\d_{\rm SS}^2= 
C^{\a} \oC^{\dot \b} \pa_{\a \dot \b}
\ee 

For this supersymmetric case, it is a little tricky  to write down the \EI s, and it is very useful to use superspace notation to construct them.
The following  expression can be used to construct the simplest \EI s for the free chiral action:

\be
 {\widehat {\phi}}_{i \dot \a}
=
 \lt [
{\widehat {\Lambda}}_{i} \oC_{\dot \a} 
+ \ovD^2 \lt (
{\widehat \A}_{i} \oq_{\dot \a} 
\rt )
\rt ]
\la{opsihat}
\ee
The expression above is chiral
\be
\ovD_{\dot \b}
{\widehat \phi}_{i \dot \a}
= 0
\ee
and it transforms like a chiral 
superfield\footnote{The complex conjugate \be
 {\widehat {\ov \phi}}^{i}_{ \a}
=
\lt [
{\widehat {\ov \Lambda}}^{\;i} C_{  \a} 
+ D^2 \lt (
{\widehat A}^{i} \q_{  \a} 
\rt )
\rt ]
\la{psihat}\ee
transforms like an antichiral superfield, and all the above remarks apply {\em mutatis mutandis} to the complex conjugate.}
for the free theory. 
\be
\dF  
 {\widehat \phi}_{i \dot \a}
=
\d_{\rm SS} 
 {\widehat \phi}_{i \dot \a}
\la{poeculiarfree}
\ee
The above equation (\ref{poeculiarfree})
 is a peculiar one.
For the free theory, it equates the result of the operation of the functional derivative operator 
$\dB  $ in (\ref{bigbrstsusy}) to the result of the operation
of an ordinary derivative operator
 (\ref{ssderivative}).  In general these will not yield the same result, but for this specific combination, they do.
 However when one adds the superpotential to make an interacting theory, it has an extra term:
\be
\d_{\rm Interacting\;SUSY} 
  {\widehat \phi}_{i \dot \a}
=
\d_{\rm SS} 
 {\widehat \phi}_{i \dot \a}
 + g_{ijk} {\widehat A}^j {\widehat A}^k
\oC_{\dot \a}
\la{poeculiar}
\ee

For the interacting theory, we need to incorporate the above expression into a term with two fields multiplied together, as we did for the gauge theory 
in \ci{YMextrainv}.  As it did in \ci{YMextrainv}, this procedure enables us  to constuct a more complicated \EI\ from the most basic one.  So define 

\be
 {\widehat \Phi}_{ T,\dot \a} 
= T^i_j
{\widehat A}^j
{\widehat {\phi}}_{i \dot \a}
= T^i_j
{\widehat A}^j
\lt [
{\widehat {\Lambda}}_{i} \oC_{\dot \a} 
+ \ovD^2 \lt (
{\widehat \A}_{i} \oq_{\dot \a} 
\rt )
\rt ]
\ee
where the tensor $T$ is to be determined. 
Then if the following equation is satisfied
\be
T^s_i g_{sjk}  A^i A^j A^k=0
\la{chiralconstraint1}
\ee
it follows that this composite expression will transform as if it were a superfield (though it is not one):
\be
\d_{\rm Interacting\;SUSY} 
  {\widehat \Phi}_{ T,\dot \a} 
=
\d_{\rm SS} 
 {\widehat \Phi}_{ T,\dot \a} 
\ee

Then the following expression ${\Q}_{T \dot \a}
$ is an \EI:
\be 
\Q_{T, \dot \a}
=\int d^4 x d^2 \q 
{\widehat \Phi}_{ T,\dot \a}  
=
\int d^4 x d^2 \q 
\lt \{
T^s_j
{\widehat A}^j
\lt [
{\widehat {\Lambda}}_{s} \oC_{\dot \a} 
+ \ovD^2 \lt (
{\widehat \A}_{s} \oq_{\dot \a} 
\rt )
\rt ]
\rt \}
\la{psihatint}\ee

To verify that this is in the cohomology space, one can write the expressions out in detail in components, which is not that hard, or else rely on the more abstract and general proofs in \ci{jumps}.

However, we are still not finished.  In the Yang-Mills case in \ci{YMextrainv}, we had constructed a Lorentz Scalar \EI\ at this stage, whereas here we have got a Lorentz Spinor \EI.  We cannot add the expression 
(\ref{psihatint}) to the 
Action (\ref{fullsusyaction}),
 because (\ref{psihatint}) is not a Lorentz scalar.  We  need to take another step here and couple it to a new chiral dotted spinor superfield ${\widehat \w}^{\dot \a}_{ T}  $ with the appropriate quantum numbers and spin. 
Then we have a \EI\ which is a Lorentz scalar: 
\be 
{\cal E}_{T}
=\int d^4 x d^2 \q 
{\widehat \Phi}_{ T,\dot \a}  
{\widehat \w}^{\dot \a}_{ T}  
\la{psihatintscalar}
\ee
and this can be added to the action, and we can proceed to complete the Poisson Bracket as we did in
 \ci{YMextrainv}, except that we need to add transformations and an action for the introduced field ${\widehat \w}^{\dot \a}_{ T}$.  Also we need to find the BRST cohomology of the fields ${\widehat \w}^{\dot \a}_{ T}$ to understand better what happens when they are added.  That is a big topic which we shall not try to discuss here, except to say that there are plenty of \EI s there too.

How does one find solutions to the symmetrization problem
 (\ref{chiralconstraint1})?
One clue to finding a non-trivial  solution of this equation  is provided by the discussion of gauge fields  in \ci{YMextrainv}. The relevant term 
in \ci{YMextrainv} for these purposes is
\be
{\cal A}_{\rm Trilinear \; Symmetric}=
2\int d^4 x \;
\pa_{\m} A_{\n}^a
A^{\m b} A^{\n c}
f^{abc}
\la{trilineargauge1}
\ee
The reason (\ref{trilineargauge1})  does not vanish is that it is  antisymmetric in two different sets of indices, so that it can be symmetric overall, as is necessary since the fields $A_{\m}^{ a}$ are commuting bosonic fields\footnote{It is totally symmetric through doubled antisymmetry, as one can verify using integration by parts, using the fact that $f^{abc}$ is totally antisymmetric.}.
The term  (\ref{trilineargauge1}) is of course the starting term for the construction of the  \YM.
 Note that the gauge fields $A^a_{\m}$ there have two indices, an isopin index and a Lorentz index, and so it is possible to `{\bf doubly antisymmetrize}', as in the term
 (\ref{trilineargauge1}), without getting zero. 

The superpotential has the symmetric form:
\be
P[A] =
g_{ijk} A^i A^j A^k
\la{superpot}\ee
 Chiral superfields   have only one index--the isospin index, because they are Lorentz scalars, and that is why
(\ref{superpot}) has a symmetric form.
Looking at (\ref{trilineargauge1}) suggests we might want to get some antisymmetry into (\ref{superpot}) if we can. Then we might be able  to get a non-trivial solution for 
(\ref{chiralconstraint1}), as we did for \YM.

In fact it is easy to get two indices into each chiral superfield by simply using  chiral superfields with {\bf  two isospin-type indices}. Then one can {\bf doubly antisymmetrize} without getting zero for (\ref{superpot}).
Then there is a chance of satisfying  equations like
  (\ref{chiralconstraint1}) while having  (\ref{superpot}) nonzero. 
The most obvious way to introduce doubled antisymmetry would be to take the following expression:
\be
P[A] = g_{MNP} A^M A^N A^P\equiv
\ve_{IJK} \ve_{ABC} A^{IA} A^{JB} A^{KC}
\la{symtensormultsymmofsm23}
\ee
where the superfields now have two indices
$A^{IA}$ and the  indices must, at the very least, take three different values $A,B,C=1,2,3$ and $I,J,K=1,2,3$, 
so that this expression
 (\ref{symtensormultsymmofsm23}) does not vanish. 
 This is of course the same as taking the Determinant for this matrix.   And as we have noted in section
 \ref{introsection}, this may well be exactly what Nature has chosen for the basic SUSY Standard Model.
 The reason for the three generations of Quarks and Leptons, and for the explicit breaking of the 
$SU(3)\times SU(3)$ symmetry, is far from clear   at this stage, but at least they do not spoil this feature. 

One can also turn the constraint equation around as follows.  Define
the following differential operator
\be
{\cal L}_{\rm T} =T^i_j
A^j \fr{\pa}{\pa A^i}
\la{liederiv}
\ee
Then the equation (\ref{chiralconstraint1}) can be written in the form
\be
{\cal L}_{\rm T} P[A]=0
\ee
 
In other words, there is a solution of 
(\ref{chiralconstraint1}) for every independent operator 
(\ref{liederiv}) in the Lie algebra of invariance for the superpotential (\ref{superpot}).

So if we want to maximize the number of \EI s that can be extended to the interacting theory, while also being economical in terms of complications, we want to find a superpotential with the maximum possible invariance group for a given number of fields, while observing that the superpotential should be cubic.  Once again, this points to a Determinant, which has a very large invariance group.

However it should be noted that again the answer to the Bootstrap is not unique.  Any invariance in the superpotential will give rise to an extended \EI.  So we are still a long way from `deducing the SSM' from the free theory. On the other hand, there does appear to be something of that kind going on here.

\section{Lie Algebra of Invariance for the Basic DASSM}

\la{basicliealg}

Here we will look more closely at the $SU(3) \times SU(3)$ invariances of the Basic DASSM, and at the detailed form of the \EI s for that theory.  As a start on that, we want to find all Lie derivative solutions of:

 \be
{\cal L} P =0
\ee
for the Basic DASSM double antisymmetry above, which is equivalent to:
\be
{\cal P} =  {\rm Det} \lt (
\begin{array}{ccc}
H^i & K^i & L^i\\
s S &pP
&gJ
\\
\end{array}
\rt ) =  s K^i L_i S + p L^i H_i P +g H^i K_iJ 
\la{horvertable}
\ee
If we set the couplings $s=p=g=1$ in the above, we already know all the invariances of this polynomial from its construction.  They are the generators of the group $SU(3)_{\rm Horizontal} \times SU(3)_{\rm Vertical} $.  At first this seems tricky when the coupling constants are present.  However they are easily handled.  Here is the basic idea.
Consider the surface of a sphere described by
\be
 P_{\rm Sphere} = x^2 + y^2 + z^2 =r^2
\ee
This has a group of Lie generators that yield zero on the expression.  Here is an example:
\be
L_3= x \fr{\pa}{\pa y}
-y \fr{\pa}{\pa x}
\ee
and the invariance is described by
\be
L_{3 } P_{\rm Sphere} = 0.
\ee

If we stretch the sphere to an egg:
\be
P_{\rm Egg} = x^2 + \fr{1}{3} y^2 + z^2 =r^2
\ee
that Lie operator is still present, but it 
takes the form
\be
L_{3 \;\rm Egg}= 3 x \fr{\pa}{\pa y}
- y \fr{\pa}{\pa x}
\ee
In other words:
\be
L_{3\; \rm Egg} P_{\rm Egg} = 0.
\ee

The coupling constants in the DASSM behave the same way.
It is a fact that the kinetic terms need to be renormalized in this process.  However that does not spoil the existence of the \EI s, because they only require solutions of the equation relating to the superpotential.  The full invariance, including invariance of the kinetic terms, is not needed to build the \EI s.

Using this idea one quickly can assemble the  full set of invariances for the Basic DASSM  and we summarize them in Table  (\ref{basictable}).
Horizontal and Vertical refer to substitutions of those kinds in the matrix in equation (\ref{horvertable}).

{\normalsize
\be
\begin{tabular}{|c|c|c|c|}
\hline
\multicolumn{4}{|c|}{Invariances of the Superpotential for the Basic DASSM Model: Table (\ref{basictable}): Solutions of ${\cal L} {\cal P}=0$}
\\
\hline 
\multicolumn{4}{|c|}{ 
${\cal P}= g H^i K_iJ + s K^i L_i S +  p L^i H_i P$}
\\
\hline
Name & Lie Operator
&Name &\EI
\\
\hline
\multicolumn{4}
{|c|}{ \bf `Horizontal' Leptonic Lie Operators with ${\cal L} {\cal P}=0$}
\\
\hline
\hline
$
{\cal L}_{Y=2,L=-1} 
$
&$ p P \fr{\pa}{\pa J}-  g K^i \fr{\pa}{\pa L^i} $
&${\eta}^{}_{P\dot \a}$
&$ p P {\ov \phi}_{J \dot \a}-  g K^i {\ov \phi}_{L i\dot \a} $
\\
\hline
$
{\cal L}_{Y=-2,L=1} 
$&$ g J \fr{\pa}{\pa P}- p L^i \fr{\pa}{\pa K^i} $
&${\eta}^{}_{E\dot \a}$&$ g J {\ov \phi}_{P \dot \a}-  p L^i {\ov \phi}_{K i\dot \a} $
\\
\hline
$
{\cal L}_{Y=0,L=-1} 
$&$s S \fr{\pa}{\pa J}- g H^i \fr{\pa}{\pa L^i} $
&${\eta}^{}_{S\dot \a}$&$ s S {\ov \phi}_{J \dot \a}-  g H^i {\ov \phi}_{L i\dot \a} $
\\
\hline
$
{\cal L}_{Y=0,L=1} 
$&$ g J \fr{\pa}{\pa S}- s L^i \fr{\pa}{\pa H^i} $
&${\eta}^{}_{N\dot \a}$&$ g J {\ov \phi}_{S \dot \a}-  s L^i {\ov \phi}_{H i\dot \a} $
\\
\hline
\multicolumn{4}
{|c|}{ \bf `Horizontal' Non-Leptonic Lie Operators with ${\cal L} {\cal P}=0$}
\\
\hline
$
{\cal L}_{Y=2,L=0} 
$&$ p P \fr{\pa}{\pa S}- s K^i \fr{\pa}{\pa H^i} $
&${\eta}^{}_{Y=2,\dot \a}$&$ p P {\ov \phi}_{S \dot \a}-  s K^i {\ov \phi}_{H i\dot \a} $
\\
\hline
$
{\cal L}_{Y=-2,L=0} 
$&$ sS \fr{\pa}{\pa P}- p H^i \fr{\pa}{\pa K^i} $
&${\eta}^{}_{Y=-2,\dot \a}$&$ s S {\ov \phi}_{P \dot \a}-  p H^i {\ov \phi}_{K i\dot \a} $
\\
\hline
${\cal L}_{H3} 
$&$
 J \fr{\pa}{\pa J}- S \fr{\pa}{\pa S}
$&${\eta}^{\dot \a}_{3}$
&$ J {\ov \phi}_{J \dot \a}-  S {\ov \phi}_{S \dot \a} $
\\
$ 
$&$
+  L^i \fr{\pa}{\pa L^i} 
-  H^i \fr{\pa}{\pa H^i} 
$&
&$+ L^i {\ov \phi}_{L i \dot \a}-   H^i {\ov \phi}_{H i\dot \a} $
\\
\hline
${\cal L}_{H8} 
$&$
 J \fr{\pa}{\pa J}+S \fr{\pa}{\pa S}
- 2 P \fr{\pa}{\pa P}
$&${\eta}^{\dot \a}_{8}$&etc.\\&$+  L^i \fr{\pa}{\pa L^i} 
+  H^i \fr{\pa}{\pa H^i} 
-  2 K^i \fr{\pa}{\pa K^i} 
$&&
\\
\hline
\hline
\multicolumn{4}
{|c|}{ \bf `Vertical' Leptonic Lie Operators with ${\cal L} {\cal P}=0$}
\\
\hline
\hline
$
{\cal L}_{i,Y=1,L=-1}$&$ g J \fr{\pa}{\pa L^i}- sS\fr{\pa}{\pa  H^i }- p P \fr{\pa}{\pa K^i} 
$
&${\eta}^{}_{Ri \dot \a}$&$g J {\ov \phi}_{L i \dot \a}-  sS {\ov \phi}_{H i\dot \a} - p P {\ov \phi}_{K i\dot \a} $
\\
\hline
$
{\cal L}^i_{Y=-1,L=1} $&$ \fr{1}{g} L^i \fr{\pa}{\pa J}- \fr{1}{s} H^i\fr{\pa}{\pa  S }-\fr{1}{p}  K^i \fr{\pa}{\pa P} 
$&${\eta}^{i }_{L \dot \a}$
&$ \fr{1}{g} L^i {\ov \phi}_{J \dot \a}-  \fr{1}{s} H^i {\ov \phi}_{S \dot \a}-  \fr{1}{p} K^i {\ov \phi}_{P \dot \a}$
\\
\hline
\multicolumn{4}
{|c|}{ \bf `Vertical' Non-Leptonic Lie Operators with ${\cal L} {\cal P}=0$}
\\
\hline
$
{\cal L}^a 
$&$  
\s^{aj}_{i} \lt ( L^i \fr{\pa}{\pa L^j} +
 K^i \fr{\pa}{\pa K^j} + H^i \fr{\pa}{\pa H^j} 
\rt )$&${\eta}^{a}_{\dot \a}$&etc.
\\
\hline
$
{\cal L}_{V8} 
$&$ 
L^i \fr{\pa}{\pa L^i} +
 K^i \fr{\pa}{\pa K^i} + H^i \fr{\pa}{\pa H^i} 
$&${\eta}^{\dot \a}_{V8}$&etc.\\&$
- 2 J \fr{\pa}{\pa J} 
- 2 P \fr{\pa}{\pa P} 
- 2 S \fr{\pa}{\pa S} 
$& &
\\
\hline
\hline
\end{tabular}
\\
\la{basictable}
\ee
} 

        In Table (\ref{basictable}), we list the Lie Operators that comprise the generators of the $SU(3) \times SU(3)$ invariance discussed in section
 \ref{introsection} above, expressed in terms of the chiral superfields of the Basic DASSM.  From these, we then write down the related \EI s using the discussion in section
 \ref{bootdassm} above.  The next step would be to couple these \EI s to new chiral dotted superfields like ${\widehat \w}_{E}^{\dot \a} 
$ in terms like
\be 
\int d^4 x d^2 \q 
{\widehat \w}_{E}^{\dot \a} 
{\widehat {\eta}}^{}_{P\dot \a}
=
\int d^4 x d^2 \q 
{\widehat \w}_{E}^{\dot \a} 
\lt \{ 
 p {\widehat P} {\widehat {\ov \phi}}_{J \dot \a}-  g {\widehat K}^i 
{\widehat {\ov \phi}}_{L i\dot \a} 
\rt\}
\eb=
\int d^4 x d^2 \q 
{\widehat \w}_{E}^{\dot \a} 
\lt \{ 
 p {\widehat P}
\lt (  {\widehat \Lam}_{J} \oC_{ \dot \a} -\ovD^2 ({\widehat\oJ} \oq_{\dot \a})\rt )
-  g {\widehat K}^i 
\lt (  {\widehat \Lam}_{L i} \oC_{ \dot \a} -\ovD^2 (
{\widehat\oL}_i \oq_{\dot \a})\rt )
\rt\}
\la{exampleofaddedterm}
\ee
and then complete the action by adding terms needed to satisfy the relevant \PB\ in full \ci{susybreaks,canjphys}.

Observe that the above contains a term that can be written as an integral over all superspace:
\be
\int d^4 x d^4 \q 
\lt \{
{\widehat \w}_{E}^{\dot \a} 
g {\widehat K}^i 
 {\widehat \oL}_i \oq_{\dot \a} 
\rt \}
\la{couplingterm}
\ee
This  
gives rise to a mass mixing term when the Higgs field
gets a VEV: 
\be
\lt < {\widehat K}^i\rt > \ra 
m   {k}^i 
\ee

  The only way that these kinds of expressions could be interesting is if they generate a sensible mass spectrum. That seems to be the most important question.
It is clear that we need to add more
to this theory to answer it, because some of the vector boson leptons that arise from the simplest additions are clearly tachyonic \ci{susybreaks}. This can possibly be cured by continuing to add more \EI s, and there are an infinite number available as can be seen from 
\ci{jumps}. Observe however, that the Leptonic terms in the Table  (\ref{basictable}) that will generate mass terms after gauge symmetry breaking, from terms like 
(\ref{couplingterm}), are matched nicely to the family of Leptons.  Is it possible that the muon and Tau families are actually in the coupled terms rather than in the basic chiral superfields that we start with?
Of course, even if this works, it still leaves  the Quarks in   mystery.

  The Higgs singlet J plays an important role too. If one adds the term $g' m^2 {\widehat J}$ to the DASSM superpotential 
 \ci{superfieldnotation}, then the vacuum gets a non-zero energy expectation value (`VEV') for its energy density, which can be shifted back to zero.  This VEV can be returned to zero by the  shifts $H^i\ra m h^i +H^i,K^i\ra m k^i +K^i$, with $h^i k_i + g'=0$.
This shift prevents the spontaneous breaking \ci{Oraiff} of SUSY, by renormalizing the VEV of the auxiliary field $\lt <\oF_J \rt >\ra 0$,
\be
-\lt < \fr{\pa {\cal P}}{\pa F_J}
\rt >
= -\lt < g' m^2 + H^i K_i - \oF_J 
\rt > \ra   \lt <  \oF_J 
\rt > 
\ra 0
\ee so that the action has zero ground state energy. 
This also gives masses to the Quarks and Leptons and some Gauge/Higgs particles, and it breaks the gauge symmetry $SU(2) \times U(1) \ra U(1)$.

However, the term $g' m^2 {\widehat J}$  is not an invariant of the  $SU(3) \times SU(3)$ symmetry of the action (\ref{simplest}), or its version with three generations in
 (\ref{bigpotential}).  If one adds the term $g' m^2 {\widehat J}$,   and also adds terms like
 	(\ref{exampleofaddedterm}), whose supersymmetry depends on the invariance of the superpotential, supersymmetry is explicitly broken in a very specific way that vanishes when the mass vanishes.

To unravel the consequences of the resulting mass mixing requires a knowledge of the BRST cohomology of the dotted chiral spinor superfields.

As can be seen from Table
 (\ref{basictable}), if  S and J are not present in the theory, then much of the above simply does not happen.  The completion of the matrix $A^{AI}$ with these is essential to the construction of these \EI s.

\section{The Full \DS}

\la{fulldassmsec}
To incorporate all the three generations of Leptons and Quarks one could take:
\be
A^{IA} \equiv \lt (
\begin{array}{ccc}
H^i & K^i & L^i\\
s S & p P
&g J
\\
\end{array}
\rt )
\ra
\lt (
\begin{array}{ccc}
H^i & K^i & L^{ip} \oplus Q^{ipc}\\
s_{pq} S^{q} \oplus t_{pq} T^{q}_{c} &p_{pq} P^{q} \oplus b_{pq} B^{q}_{c} 
&g J
\\
\end{array}
\rt )
\la{bigfatmatrix}
\ee
and then one gets the following superpotential, which is closely related to  (\ref{basicPfromdet}):
\be
{\cal P}= g H^i K_iJ 
+ K^i \lt ( L_{i}^{p} s_{pq} S^q + Q_{i}^{c p} t_{pq}
 T^q_c  \rt )
+  \lt (
L^{i p} p_{pq} P^q + Q^{i c p} b_{pq}
 B^q_c  \rt )H_i
\la{bigpotential}
\ee 

This form is still doubly antisymmetric (roughly speaking) in its terms, and it still has lots of invariance, though it is not so obviously the determinant of any simple matrix.  The quantum numbers of these fields are summarized in Table ({\ref{cssmtable}).

{\small
\be
\begin{tabular}{|c|c|c
|c|c|c
|c|c|c|}
\hline
\multicolumn{8}
{|c|}{Table (\ref{cssmtable}): The Chiral Superfields in the DASSM
}
\\
\hline
\multicolumn{8}
{|c|}{ \bf   Left Weak Doublet Matter 
SuperFields}
\\
\hline
{\rm Field} & Y 
& {\rm SU(3)} 
& {\rm SU(2)} 
& {\rm F} 
& {\rm B} 
& {\rm L} 
& {\rm D} 
\\
\hline
$ L^{pi} $& -1 
& 1 & 2 
& 3
& 0
& 1
& 1
\\
\hline
$ Q^{cpi} $ & $\fr{1}{3}$ 
& 
3 &
2 &
3 &
 $\fr{1}{3}$
& 0
& 1
\\\hline
\multicolumn{8}
{|c|}{ \bf  Right Weak Singlet Matter 
SuperFields}
\\
\hline
$P^{ p}$ & 2 
& 1
& 1
& 3
& 0
& -1
& 1
\\
\hline
$S^{ p}$ & 0 
& 1
& 1
& 3
& 0
& -1
& 1
\\

\hline
$T_c^{ p}$ & $-\fr{4}{3}$ 
& ${\ov 3}$ &
1 &
3 &
 $-\fr{1}{3}$
& 0
& 1
\\
\hline
$B_c^{ p}$ & $\fr{2}{3}$ 
& ${\ov 3}$ 
&
1 &
3 &
 $- \fr{1}{3}$
& 0
& 1
\\
\hline
\multicolumn{8}
{|c|}{ \bf Higgs SuperFields}
\\
\hline
$J$
& 0 
& 1
& 1
& 1
& 0
& 0
& 1
\\
\hline
$H^i$ 
& -1 
& 1
& 2
& 1
& 0
& 0
& 1
\\
\hline
$K^i$ 
& 1 
& 1
& 2
& 1
& 0
& 0
& 1
\\
\hline
\end{tabular}
\\
\la{cssmtable}
\ee
\normalsize
}
In Table (\ref{cssmtable}), Y is weak hypercharge, F stands for the number of families for each superfield, B is baryon number, L is Lepton number and D stands for mass dimension. After spontaneous gauge symmetry breaking of 
$SU(2)_{\rm Weak}\times U(1) \ra U(1)$ the left hand doublets break into 
\be
L^{pi} \ra \lt ( 
\begin{tabular}{c}
$N^p$ \\$E^p$
\end{tabular}
\rt ), 
 Q^{cpi} \ra \lt ( 
\begin{tabular}{c}
$U^{cp}$\\$D^{cp}$
\end{tabular}
\rt )
\ee

The notation for these left doublet and right singlet Leptons and Quarks is designed to prevent the need for repetition of the notation for the field and its conjugate, which would create even more indices on the already index-encrusted fields. 

As we noted above, this might be the wrong approach, since the theory already has enough room to make the families without adding these (except for the mystery of why there are Quarks).

\section{Conclusion}
\la{vevsection}

So does SUSY `know' about the Standard Model, in the sense that the addition of the \ELI s require us to start with the DASSM?  SUSY appears to be somewhat aware of it, because if one wants a maximal solution of the constraints with a minimal number of superfields, one gets the simple action that generates the DASSM.  

However there are several steps involved, and the derivation of  \YM\ from the unique \ELI\ in free gauge theory in \ci{YMextrainv}  is more straightforward, and much simpler, than what happens for the SUSY theory. However, if we decide to add fields so that the maximum number of \ELI s of SUSY can be added to the action in the simplest way, we do get the DASSM.

We have looked at the extension of
 \EI s, starting with free gauge theory and with free chiral supersymmetry.  With some reasonable assumptions, these yield, respectively, Yang-Mills gauge theory and a special, and very simple, version of the SSM, namely the basic DASSM.  There is a parallel between the construction of Yang-Mills theory  and the construction of the basic DASSM  from the free theories.   In particular, both  constructions use double antisymmetrization in a fundamental way as part of the extension.

From these constructions, it seems that the cohomological properties of the free theories are influencing the interacting theories that are actually chosen by Nature.  It also appears  that SUSY helps to explain the Standard Model, by showing why the left weak doublets and right weak singlets make things simpler and more natural in the matrix $A$  that yields the  simple superpotential $\rm Det \;A$. If one wants to proceed in the simplest and most natural way, there is a limited choice about how to extend the  \EI s from the free theories to the interacting ones. 

 These results  justify the term `Bootstrap'.  But it must be admitted that any Yang-Mills theory would suffice to extend the \EI\ in free gauge theory, and any invariance of the superpotential would generate an extension of the chiral SUSY \EI s. So the Bootstrap needs to be supplemented by some   principles of economy in the number of fields, simplicity in the interactions, and maximality of the number of \EI s, to derive the Basic DASSM.

Even if we accept those principles, Nature adds some very non-minimal complications to the Basic DASSM,   in the form of the $SU(2) \times U(1)$ gauge fields, the multiple families of Leptons, the multiple families of Quarks, and the colour SU(3) gauge fields. While these are consistent with the Bootstrap construction above, they certainly go beyond the minimal solution.  Where do these come from?

What happens if this `Bootstrap procedure' is applied to SUSY Yang Mills?  It   appears that there are \EI s in SUSY Yang Mills, but the details have not been worked out yet.
Would that add more constraints that might help to account for the non-minimal structure seen in the Full Standard Model?

  The right handed Neutrino S and the singlet Higgs J
are essential to the construction of the Basic DASSM, with its simple 
action (\ref{simplest}).
The closely related action (\ref{bigpotential})  for the Full DASSM also requires flavoured right Neutrinos $S^p$ and the singlet Higgs J.
 The right handed Neutrino superfields  yield massive Neutrinos and so this model is, at least, not obviously wrong.  But the fundamental issue to be resolved is the mass spectrum we obtain from the extension of this theory.  There are certainly tachyons if one does this in the simplest way.

\vspace{.2in}

\begin{center}
 {\bf Acknowledgments}
\end{center}
\vspace{.2in}
  I thank  Carlo Becchi, Philip Candelas, Rhys 
Davies,  Paul Frampton, John Moffat,   Raymond Stora, Xerxes Tata and
J.C. Taylor  for stimulating correspondence and conversations, and the Perimeter Institute for hospitality.


\tiny
\articlenumber
\end{document}

%% file: Johnmacros15.tex
\newcommand{\YM}{Yang-Mills theory}

\newcommand{\ELI}{Extraordinary Lorentz Invariant}
\newcommand{\dF}{\d_{\rm Free}}
\newcommand{\DS}{DASSM}

\newcommand{\EI}{Extraordinary Invariant}

\newcommand{\PB}{BRST Poisson Bracket}

\newcommand{\dB}{\d_{{}_{{}_{\rm BRST}}}}

\newcommand{\SM}{Standard Model}

\newcommand{\LT}{\LaTeX}

\newcommand{\bt}{\begin{tabular}{c}}
\newcommand{\et}{\end{tabular}}

\newcommand{\eb}{\ee\be } 
 
\newcommand{\bmat}{\lt ( \begin{array} }
\newcommand{\emat}{  \end{array} \rt )}

\newcommand{\oQ}{{\ov Q}}

\newcommand{\ovD}{{\ov D}}

\newcommand{\oJ}{{\ov J}}

\newcommand{\oq}{{\ov \q}}

\newcommand{\ED}{\end{document}}

\newcommand{\og}{{\ov g}}

\newcommand{\oC}{{\ov C}}
\newcommand{\oL}{{\ov L}}

\newcommand{\oF}{{\ov F}}
\newcommand{\A}{{\ov A}}

\renewcommand{\a}{\alpha}	
\renewcommand{\b}{\beta}

\renewcommand{\d}{\delta}

\newcommand{\ve}{\varepsilon}

\newcommand{\q}{\theta}

\newcommand{\m}{\mu}
\newcommand{\n}{\nu}

\newcommand{\s}{\sigma}

\newcommand{\y}{\psi}
\newcommand{\w}{\omega}

\newcommand{\Q}{\Theta}	
\newcommand{\Lam}{\Lambda}

\newcommand{\la}{\label}
\newcommand{\ci}{\cite}

\newcommand{\ds}{\documentstyle}	
\newcommand{\fr}{\frac}

\newcommand{\pa}{\partial}
\newcommand{\ov}{\overline}
\newcommand{\be}{\begin{equation}}
\newcommand{\ee}{\end{equation}}
\newcommand{\ba}{\begin{array}} 
\newcommand{\ea}{\end{array}}
\newcommand{\bea}{\begin{eqnarray}}
\newcommand{\eea}{\end{eqnarray}}
\newcommand{\ra}{\rightarrow}

\newcommand{\lt}{\left}
\newcommand{\rt}{\right}

\newcommand{\ben}{\begin{enumerate}}
\newcommand{\een}{\end{enumerate}}
\newcommand{\bitem}{\begin{itemize}}
\newcommand{\eitem}{\end{itemize}}

%% file: 3672.DASSM.bbl
\begin{thebibliography}{99}

\bibitem{YMextrainv} J. A. Dixon,  
`\EI s are Seeds for New Theories', arXiv [HEP-TH] 
Preprint April 2013.


\bibitem{SU5} A recent article on the Unification of the Standard Model with the group SU(5) is: Gauged Flavor, Supersymmetry and Grand Unification,
Rabindra N. Mohapatra, arXiv 1205.6190v1
[hep-ph] 

\bibitem{ExptSU5} The unification into the groups SU(5)  predicts proton decay at a level that is too large.  See \ci{SMexpmt} for details.


\bibitem{SMexpmt} The experimental testing of the Standard Model has been summarized in the publications of 
 J. Beringer et al. (Particle Data Group), Phys. Rev. D86, 010001 (2012) also available on the Internet.

\bibitem{SSM} A good place to find literature about the SSM is in the various SUSY conferences, for example in citations  \ci{SUSY09}  etc.   up to the present time and also in the Particle Data Group Summaries which are available on the Internet.

\bibitem{SUSY09} SUSY 2009:  AIP Conference Proceedings Volume 1200, Boston, Massachusetts. Eds: George Alverson,
    Pran Nath, Brent Nelson. 




\bibitem{massneutrino} The various issues about neutrino mass are discussed in the Particle Data Group summary 
 Neutrino Mass, Mixing and Oscillations,
Updated  May  2012  by  K.  Nakamura  (Kavli  IPMU  (WPI),  U.  Tokyo,   KEK)   and S.T. Petcov 
(SISSA/INFN Trieste, Kavli IPMU  (WPI), U. Tokyo, Bulgarian  Academy  of Sciences).





 \bibitem{superfieldnotation}
Our notation for chiral superfields is $
{\widehat J} =
J + \q^{\a} \y_{J\a}+ \fr{1}{2}\q^{2} F_{J}
$.  Frequently in this paper we have dropped the hat $
\;{\widehat {} }
\;$ when the context was sufficient for clarity, to avoid clutter.
For SUSY textbooks, see for example 
 \ci{west}, \ci{superspace} and 
 \ci{WB}. A useful reprint collection is \ci{ferrarabook}.  

\bibitem{west}  Peter West, Introduction to Supersymmetry and Supergravity, World Scientific (1990).

\bibitem{superspace}  S. J. Gates, M. T. Grisaru, M. Rocek and W. Siegel, Superspace, Benjamin, 1983.

\bibitem{WB}J. Wess and J. Bagger,  Supersymmetry and Supergravity, Second Edition, Princeton University Press (1992).  


\bibitem{ferrarabook} Many of the original papers on  SUSY and supergravity are collected in  Supersymmetry, Vols. 1 and 2, ed. Sergio Ferrara, North Holland, World Scientific, (1987).


\bibitem{jumps} J. A. Dixon, `SUSY Jumps Out of Superspace in the Supersymmetric Standard Model', 
arXiv 1012.4773.


\bibitem{susybreaks} Ibid. The tachyonic masses for some of the Bosonic Vector particles are clear from the results in `Supersymmetry Breaks Itself for Quarks and Leptons in the SUSY Standard Model', arXiv 1012.4970. 

\bibitem{canjphys}
Ibid.  Progress on a new way to break SUSY, Can. J. Phys. to be published.

 \bibitem{Oraiff} This is the normal behaviour of pure chiral theories, as was shown by L.  O'Raifeartaigh: Nucl. Phys. B96 (1975) 331-352.   This important paper is included in Vol. 1 of \ci{ferrarabook} at p. 411.




\end{thebibliography}
